# Influence of thickness on structural, electrical and optical properties of $VO_2$ films grown on $TiO_2$-buffered c-$Al_2O_3$ substrates

M. E. Kutepov[1], V. E. Kaydashev[1*], D.V. Stryukov[2], A.S. Konstantinov[3], A.V. Nikolskiy[4], A.T. Kozakov[4], A.D. Morozov[5], I. K. Domaratskiy[5], S.S. Zhukov[5] and E. M. Kaidashev[1]

[1]*I. I. Vorovich Mathematics, Mechanics and Computer Science Institute, Laboratory of Nanomaterials, Southern Federal University, 200/1 Stachki Ave., 344090 Rostov-on-Don, Russia*
[2]*Federal Research Centre The Southern Scientific Centre of the Russian Academy of Sciences, Chekhov Ave., 41, 344006, Rostov-on-Don, Russia*
[3]*Physics Faculty, Southern Federal University, 5 Zorge St., 344090 Rostov-on-Don, Russia*
[4]*Institute of Physics, Southern Federal University, 194 Stachki Ave., 344090 Rostov-on-Don, Russia*
[5]*Moscow Institute of Physics and Technology (MIPT), Institutskiy 9, 141701 Dolgoprudny, Russia*

Corresponding author: kaydashev@gmail.com*

**Abstract**

Vanadium dioxide with metal-to-insulator transition (MIT) that is triggered by heat, current or light is a promising material for modern active THz/mid-IR metasurfaces and all-optical big data processing systems. Multilayer $VO_2$-based active metasurfaces are urgently needed however several important issues related to $VO_2$ properties in $VO_2/TiO_2/Al_2O_3$ films should be thoroughly examined first. We study electrical, optical and structural properties of $VO_2$ films as well as their composition and switching characteristics as function of the $VO_2$ layer thickness in $VO_2/TiO_2$ composites. XRD analysis revealed an epitaxial growth of films with deformation of the monoclinic $VO_2$ lattice to hexagonal symmetry. Reduced $VO_2$ layer thickness from 170 nm to 20 nm results in increased phase transition temperature while the width of the resistance versus temperature hysteresis loop *R(T)* remains constant at ~6 °C for all $VO_2$ thicknesses in the range of 20-170 nm. The resistance alteration ratio $\Delta R_{ratio}$ is reduced from $4{,}2\times10^3$ to $2.7\times10^2$ in thinner films. Raman spectra reveal a significant shift of $VO_2$ lattice vibration modes for films thinner than 30 nm claiming a great structural strain whereas modes position for thicker VO2 layers are similar to those in bulk structure. Composition of $VO_2$ films has revealed only a minor alteration of $VO_2/V_2O_5$ phases ratio from 1.6 to 1.8 when the film thickness has been increased from 20 nm to 50 nm. Investigation of surface elemental composition and valence states of $VO_2$ films revealed that $VO_2/V_2O_5$ ratio remains practically unchanged with thickness reduction. The study of electrical MIT dynamics revealed the switching time of a 50 nm VO2 film to be as low as 800 ns.

## 1. Introduction

Vanadium dioxide has recently attracted much attention as a prospective material for active THz/ middle IR metasurfaces [1–5], NIR waveguide modulators [6–8], LIDARs [7] and big data development using all-optical image processing [9], middle IR photodetectors [10, 11], smart windows [12], and other applications in optoelectronics.

High quality epitaxial $VO_2$ films do typically show resistance change for 3-4 orders of magnitude due to metal-to-insulator transition (MIT) as well as narrow R(T) hysteresis loop which are both key important properties for fast and deep optical modulation in all near IR, middle IR and THz ranges.

Historically first, high crystalline epitaxial films were grown on sapphire or titanium dioxide monocrystal substrates, which best fit to lattice parameters of vanadium dioxide [13]. Later, many efforts were undertaken to prepare $VO_2$ films with needed electric characteristics on cheaper silicon [7, 8], $TiO_2$-buffered $SiO_2$/Si and flexible $TiO_2$/polyimide substrates [14]. However, the former $VO_2$ films are typically amorphous and do reveal a resistance alteration of only 1.5-2 orders of magnitude due to MIT.

Multilayer structures with high quality epitaxial $VO_2$ layers that possess simultaneously excellent structural characteristics, large IR/THz modulation, narrow resistance hysteresis loop and fast electric switching are highly desired. In particular, such films are expected to boost the progress in multi-layer $VO_2$-based active THz/IR metasurfaces the technology of which is still challenging. $VO_2$-based active metasurfaces should be cost-effective and, thus, might be controlled with direct thermal heating or indirect heating by microsecond/nanosecond electric current pulses or by cheap continuous lasers. In all these cases a metal-to-isolator transition is occurred in $VO_2$ due to temperature induced structural transition in a time window from microseconds [7] to seconds and shorter responses of several tens of nanoseconds are rear reported [6]. Indeed, $VO_2$ device backward switching rate is limited by the process of the heat dissipation that is function of a film thickness. The reduction of $VO_2$ layer thickness allows one boost the backward $VO_2$ switching rate [15]. However, when the thickness of the $VO_2$ layer is only 10 nm or less, typically a significant lattice strains are observed. This results in a significant degradation of the *R(T)* hysteresis loop, namely, a broadening as well as a decreased ratio of films conductivity in metallic and dielectric states [16].

Several efforts were undertaken to follow the influence of substrate type and $VO_2$ thickness on the electric switching dynamics. Namely, $VO_2$ films grown on expensive single crystalline $TiO_2$ substrates [17] and more cost-effective $TiO_2$-buffered $Al_2O_3$ substrates [18] were characterized. Remarkable, that $VO_2$ films prepared on single crystalline $TiO_2$ do reveal only a bit less strains compared to $VO_2$ layers grown on much more cost-effective $Al_2O_3$ substrates [17]. Also, structural strains become more pronounced in ultrathin $VO_2$ films prepared on $TiO_2$/m-$Al_2O_3$ substrates which results in significant shift of MIT hysteresis loop to room temperature [19].

The forward and backward switching times more often were studied for Si near IR waveguide modulators [6–8] and best backward switching times as short as ~36 ns and ~74 ns were achieved [6]. However, it is not clear from these studies either these times correspond to complete switching between semiconductor and metallic states or just to a minor resistance alteration which is typically occurred faster. Indeed, even partial electric switching that is occurred in first quarter of MIT hysteresis is already capable to result in significant IR waveguide modulation [7]. However, the minor resistance alteration is not enough to modulate middle IR and THz reflection/transmission. Moreover, very controversial results obtained in [19-21] and in [9] on the influence of structural strain in $VO_2$ lattice on the optical and electric switching characteristics need further experimental efforts to shed more light on this topic.

Meanwhile, it is already now clear that thin $VO_2$ films on $TiO_2$-buffered c-$Al_2O_3$ substrates are promising candidate to fabricate cost-effective quality multilayer active THz/middle IR metasurfaces. Such films are the subject of the present careful examination. First, the use of a $TiO_2$ sub-layer is dictated by the need to replace an expensive single-crystalline $TiO_2$ substrate by more cost-effective counterpart. Indeed, quite good fit of lattice parameters at both interfaces of $VO_2$/$TiO_2$ and $TiO_2$/$Al_2O_3$ makes it possible to obtain high quality epitaxial layers [20]. Second, the technology for multilayer $VO_2$/$TiO_2$/$VO_2$/$TiO_2$/…/$Al_2O_3$ structures is urgently needed. Indeed, $TiO_2$ layers act also as dielectric spacers between active $VO_2$ layers and combining functional and dielectric layers is beneficiary. However, there is still a trade-off between the film thickness, crystalline structure quality, reverse switching rate and a conductivity modulation depth.

The $TiO_2$ buffer layer deposition temperature in the range of 650-700 °C as well as $TiO_2$ sub-layer thickness were both found to influence the abruptness and position of the MIT hysteresis loop of $VO_2$ [16]. Also, they have found that the $VO_2/TiO_2$/m-$Al_2O_3$ structure is already nearly completely unstrained and a MIT is observed at ~50 °C when the thickness of a $TiO_2$ buffer layer is greater than ~200 nm [16].

We study optimal pulsed laser deposition regimes to fabricate epitaxial $VO_2$ films on $TiO_2$-buffered c-$Al_2O_3$ substrates that will be further used in multi-layered active THz and middle IR metasurfaces. To assemble more complete picture on properties of $VO_2/TiO_2$/c-$Al_2O_3$ composites we study the films obtained at varied oxygen pressure and $VO_2$ layer thickness. Namely, we analyze films structural properties, phase composition, the influence of temperature triggered MIT on conductivity, middle IR reflection and lattice vibration dynamics. Electrically induced switching dynamics is examined as well.

## 2. Experimental

First, we studied the properties of $VO_2$ films prepared at $TiO_2$/c-$Al_2O_3$ substrates at varied oxygen pressure. For this purpose, a series of $VO_2$ films was synthesized using the PLD method on c-$Al_2O_3$ substrates with preliminary deposited $TiO_2$ buffer sub-layer. The $VO_2$ ceramic target was positioned at distance of 5 cm from the substrate. Radiation of KrF laser (CL7100, 248 nm, 10 Hz) was focused onto the surface of ceramic rotating target to obtain fluence of 2 J/cm$^2$. The substrate temperature was maintained at 550°C, and the oxygen pressure was altered in the range of 1 – $3\times10^{-2}$ mbar for different regimes. The rate of $VO_2$ film growing was estimated to be ~0.425 A/pulse and a ~170 nm thick film was grown for 4000 laser pulses. As prepared $VO_2$ films were *in-situ* annealed during 15 min at oxygen pressure and temperature used for synthesis. The $TiO_2$ buffer layers were deposited at temperature of 650 °C and oxygen pressure of $3\times10^{-2}$ mbar, respectively. The ~170 nm thick $TiO_2$ layers were deposited for all samples in present study. Other parameters during the growth of $TiO_2$ sub-layer were the similar to those used for deposition of $VO_2$. Also, $VO_2$ film on base c-$Al_2O_3$ substrate without a $TiO_2$ sub-layer was deposited as a Reference Sample at oxygen pressure of $2\times10^{-2}$ mbar. Other parameters were the similar to those for $VO_2/TiO_2$/c-$Al_2O_3$ films.

Next, correlation between $VO_2$ layer thickness and the composite $VO_2/TiO_2$/c-$Al_2O_3$ properties was examined. More specifically, $VO_2$ layers with thickness of 170 nm, 50 nm, 30 nm, 20 nm we deposited on $TiO_2$-buffered c-$Al_2O_3$ substrates at oxygen pressure of $3\times10^{-2}$ mbar and temperature of 550°C.

X-ray diffraction studies were carried out using the RIKOR multifunctional X-ray complex, Bragg-Brentano focusing, Cu$K_\alpha$ radiation (30 kV, 10 mA), Ni β-filter. The scanning step was 0.01° for θ-2θ scan and 0.4° for φ scan, correspondingly.

Raman spectra of the $VO_2/TiO_2$ films were studied using Renishaw inVia Reflex Raman spectrometer with spectral resolution better than 1 cm$^{-1}$. The samples were excited at 514 nm with Ar+ laser light. A ×50 objective with a long focal distance (NA=0.5) was used for excitation and scattered light collection. A backward scattered radiation was collected in geometry, polarizers for incident and scattered light were not used. The incident light fluence did not exceed $0.74\times10^5$ W/cm$^2$ and, thus, the sample heating was minor.

X-ray Photoelectron Spectra (XPS) were obtained using ESCALAB 250 X-ray photoelectron microprobe. The samples were excited by monochromatic Al$K_\alpha$ radiation focused to the spot of 500 μm. The absolute resolvable energy interval was 0.6 eV, which was determined using the $Ag3d_{5/2}$ line. The binding energies in spectra were determined using the C1s calibration line at 285.0 eV.

Middle IR reflection spectra were recorded using Fourier Transform IR spectroscopy with Bruker Vertex 80V spectrometer and Hyperion 2000 IR microscope in ~40 μm spot.

## 3. Results and discussion

### 3.1 XRD study

The studied θ-2θ X-ray diffraction patterns revealed only reflections from the $VO_2$, $TiO_2$ layers and the $Al_2O_3$ substrate, no reflections associated with impurity phases were detected  as shown in Fig. 1 (a). The monoclinic symmetry of the unit cell with parameters a = 5.752, b = 4.538, c = 5.382 Å, and β = 122.65° is the most probable for vanadium dioxide at room temperature. In accordance with this symmetry, Miller indices are denoted below. The fact that θ-2θ X-ray diffraction patterns reveal only reflections of the (0k0) type for the $VO_2$ film and (00l) for the $Al_2O_3$ substrate indicates the co-direction of the [010]$VO_2$ and [0001]$Al_2O_3$ axes. Excellent structural properties of the films are confirmed by narrow lines in θ-2θ X-ray diffraction patterns as well as in the rocking curves as shown in Fig. 1 (b). $VO_2$ films revealed [010] axis normal to the substrate with misorientation smaller than 0.5° for all $VO_2/TiO_2/Al_2O_3$ samples. Remarkable, that for $VO_2$ layer deposited directly on $Al_2O_3$ the misorientation showed a bit higher value of 0.65°.

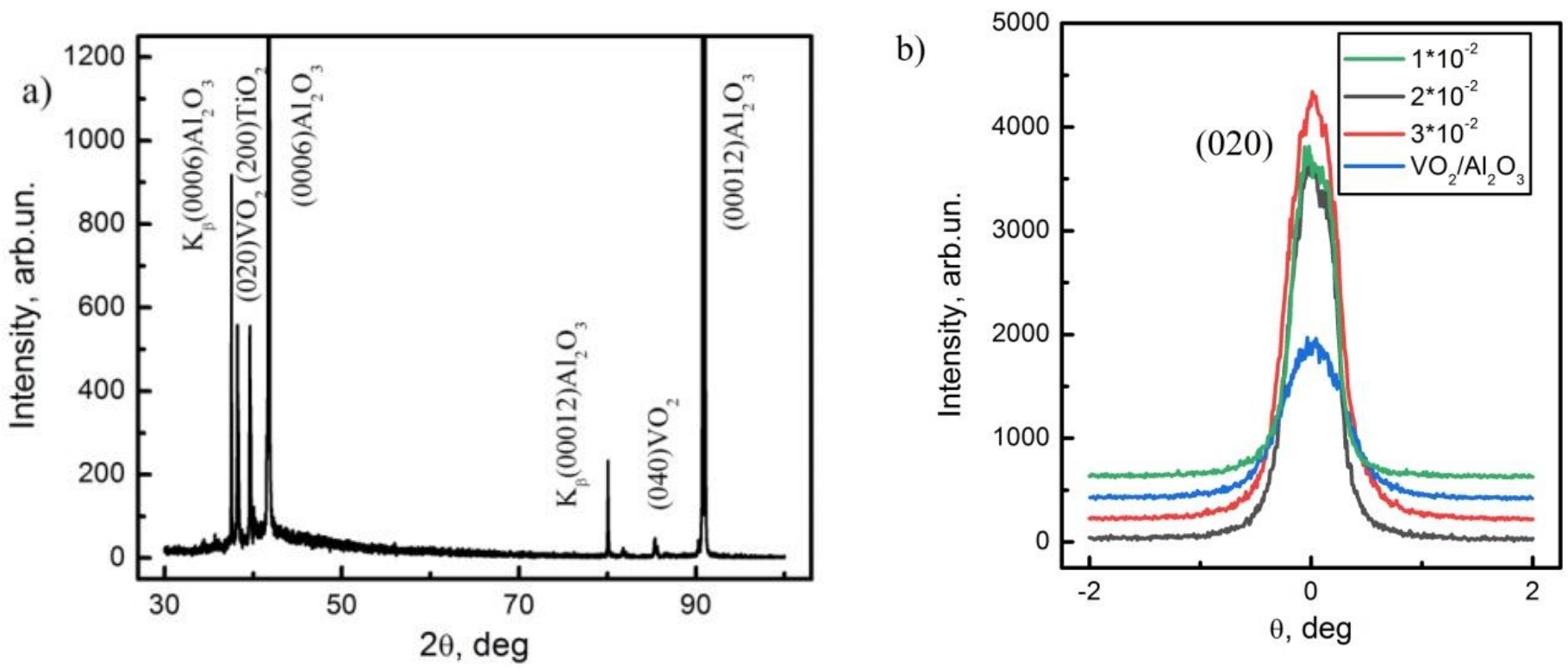

**Fig. 1** a) θ-2θ X-ray diffraction pattern for a $VO_2/TiO_2/Al_2O_3$(0006) film synthesized at $P(O_2)=1\times10^{-2}$. b) Rocking curves of the $VO_2$ (020) reflection for $VO_2$ films obtained at oxygen pressure in $1\times10^{-2}$ – $3\times10^{-2}$ range.

The epitaxial growth of all studied films was confirmed by φ-scanning of the (220) reflection as shown in Fig. 2a. More specifically, six maxima are observed on the φ-scans of (220) reflection, where the distance between adjacent maxima was measured to be 60°. This suggests that the angle β in the conjugation plane is equal to 120°. It is highly likely that the deposition of $VO_2$ on $Al_2O_3$ or on $TiO_2/Al_2O_3$ is occurred with distortion of the $VO_2$ unit cell to hexagonal symmetry for all studied films and the [010] axis of the monoclinic cell of $VO_2$ becomes a sixth order axis. Locations of the atoms of the $VO_2$ film and the $Al_2O_3$ substrate in the conjugation plane are shown in Fig. 1b and Fig. 1c, correspondingly.

The angular positions of $VO_2$ (220) reflections on φ-scans in respect to (104) reflections of the c-$Al_2O_3$ substrate indicate that $VO_2$ unit cell is a rotated for 30° in respect to the $Al_2O_3$ unit cell (Fig. 2a). From θ-2θ X-ray diffraction patterns of (0k0) $VO_2$ reflections we determine parameter "b" of the unit cell in the direction normal to the substrate surface (Fig. 2d). The "b" parameters for all films are slightly altered from 4.546 ± 0.005 Å for a film deposited on $TiO_2/Al_2O_3$ at $P(O_2)=1\times10^{-2}$ mbar to 4.517 ± 0.005 Å for a film deposited on $Al_2O_3$ without a sublayer. The maximum difference in the parameters of the unit cell "b" is 0.029 Å, which corresponds to a cell deformation not exceeding 0.6%.

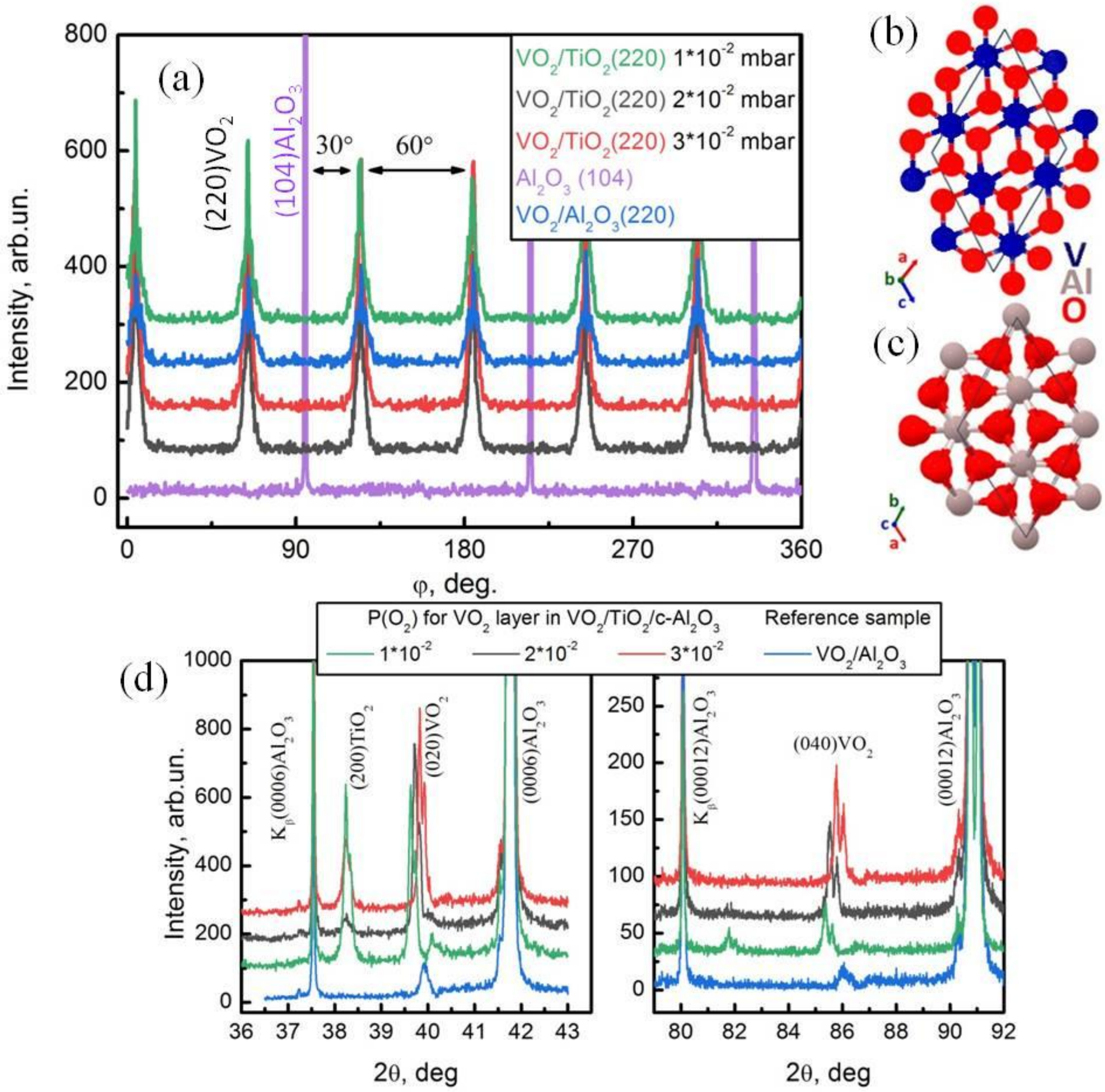

**Fig. 2** φ-scanning x-ray diffraction patterns of (220) $VO_2$ reflection of studied films and reflection (104) of the $Al_2O_3$ substrate (a). Orientation of the monoclinic $VO_2$ lattice (b) and the hexagonal $Al_2O_3$ lattice (c). Precision θ-2θ X-ray diffraction patterns of studied films (d). Blue lines in (c) and (d) correspond to $VO_2$/c-$Al_2O_3$ Ref. sample.

### 3.2 Study of resistance as function of temperature for films grown in different conditions

The MIT related alteration of the $VO_2$ layer resistance as function of temperature *R(T)* for (170 nm)$VO_2/TiO_2$/c-$Al_2O_3$ films prepared at different oxygen pressure is shown in Fig. 3. The derivative of $d(\log_{10}(R))/dT$ with minimum at MIT curve inflection point corresponding to sample deposited at oxygen pressure of $3\times10^{-2}$ mbar is shown in the inset. All the samples prepared at oxygen pressure in the range of $1\text{-}3\times10^{-2}$ mbar reveal the resistance alteration for more than 3 orders of magnitude due to MIT and small hysteresis width of 5 – 6,1 °C . The metallic state resistance is varied from 40 to 350 Ω for samples grown at different pressure. We speculate that abrupt MIT curve behavior evidences the high structural quality of $VO_2$.

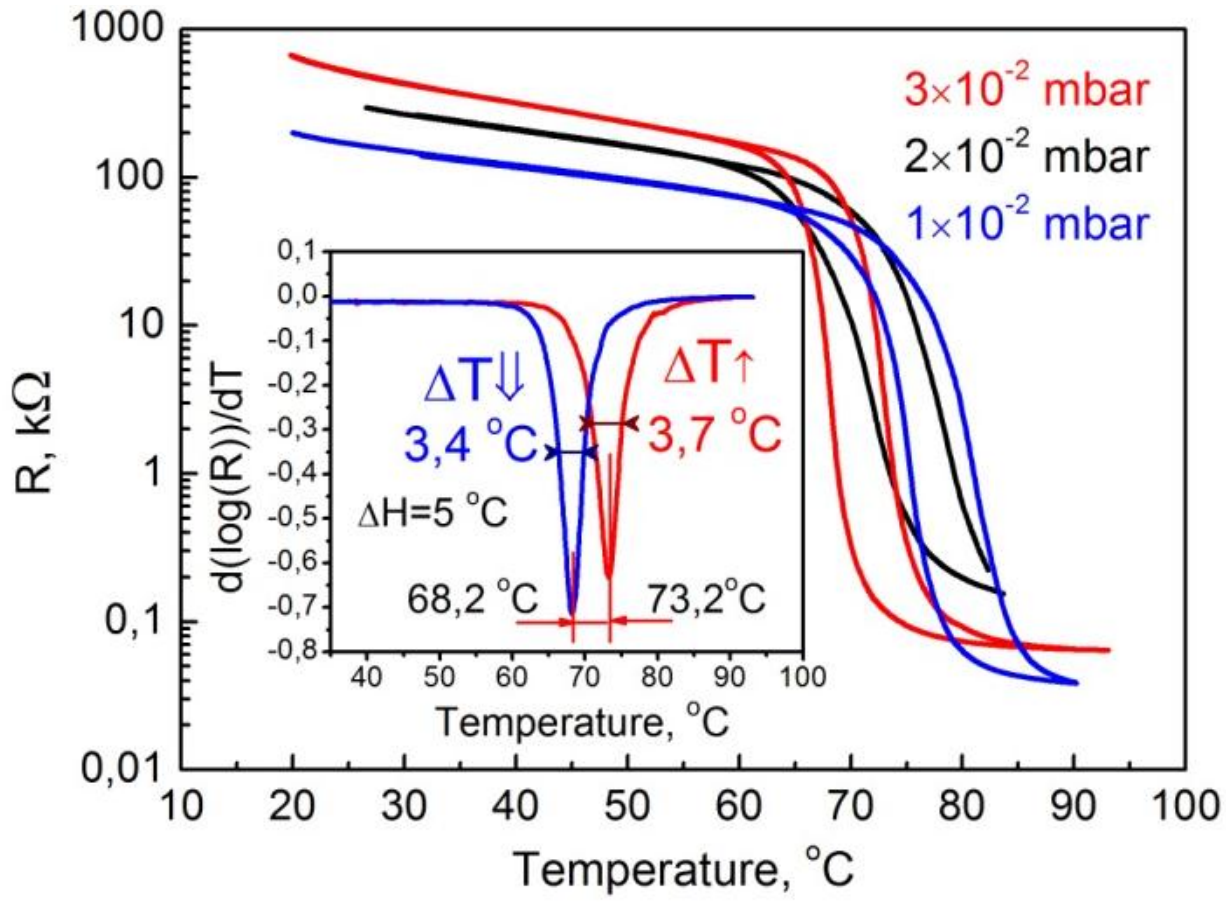

**Fig. 3** MIT induced resistance alteration as function of temperature for $VO_2/TiO_2/c\text{-}Al_2O_3$ films obtained at different oxygen pressure $P(O_2)$ in the range of 1 – $3\times10^{-2}$ mbar. The derivative of $d(\log_{10}(R))/dT$ as function of temperature for sample obtained at $P(O_2)= 3\times10^{-2}$ mbar is shown in inset.

The more detailed parameters of metal-to-insulator phase transition such as phase transition temperature ($T_{PT}$), sharpness ($\Delta T$) and width ($\Delta H$) of thermal hysteresis were defined from the temperature dependences of the derivative $d(\log_{10}(R))/dT$ shown in Fig. 4 and are summarized in Table 1.

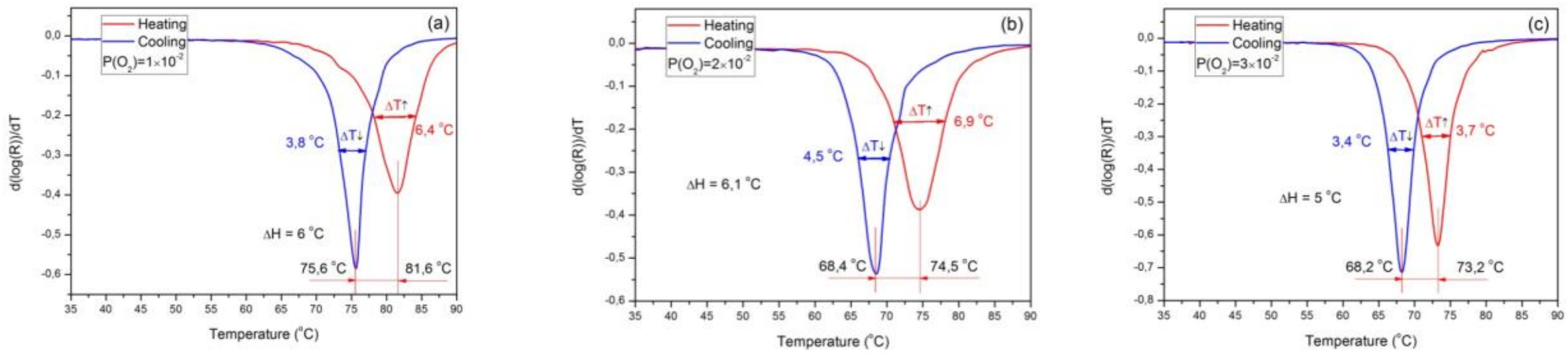

**Fig. 4** Derivative of a log(R(T)) dependence of a $VO_2/TiO_2/c\text{-}Al_2O_3$ series of samples synthesized at an oxygen pressure of $(1 – 3)\times10^{-2}$ for a $VO_2$ layer (a,b,c) and a $VO_2$ film.

According to data in Table 1 we conclude that the film deposited at oxygen pressure of $3\times10^{-2}$ mbar shows the best characteristics, namely the most abrupt MIT, greatest resistance alteration ratio $\Delta R_{ratio}$ and narrow hysteresis loop. Thus, we use this $VO_2$ film deposition regime for all the samples in second part of our study to characterize the influence of $VO_2$ layer thickness on the properties of $VO_2/TiO_2/c\text{-}Al_2O_3$ composite.

Next, we study the evolution of $VO_2/TiO_2$ composites properties of four structures with $VO_2$ layer thickness of 20 nm, 30 nm, 50 nm and 170 nm in samples S1-S4, correspondingly.

### 3.3 X-ray photoelectron spectroscopy study

The valence states of revealed elements and oxide phases content on the surface of studied composites were characterized using XPS method. Binding energies and relative content of the $V_{2p}$ and $O_{1s}$ components in XPS spectra of Samples S1-S4 are given in Table 2. All the spectra reveal two $V_{2p}$ related lines with binding energy of 516.4 eV and 517.8 eV which were assigned to belong to $VO_2$ and $V_2O_5$ oxide phases, respectively, as shown in Fig.5. The $VO_2/V_2O_5$ ratio is slightly increased from 1,6 to 1,8 for thicker $VO_2$ layers, however vanadium dioxide remains the prevailing phase for all films. Thus, the alteration of the elemental composition in thinner $VO_2$ layers is minor.

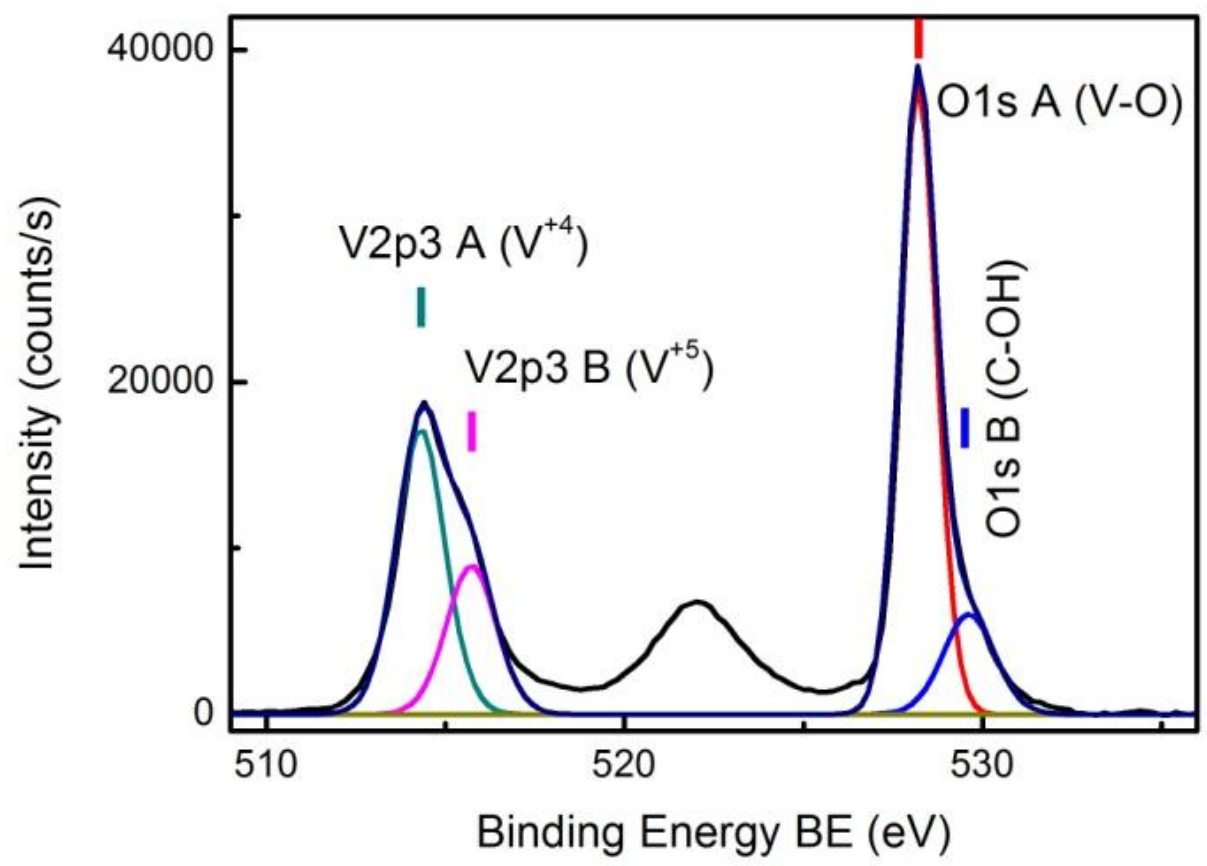

**Fig. 5** X-Ray Photoelectron Spectrum measured on the surface of $VO_2/TiO_2/c\text{-}Al_2O_3$ composite film in Sample S3.

### 3.4 Raman spectroscopy study

Raman spectra measured for samples S1-S4 with altered $VO_2$ film thicknesses are shown in Fig. 6. The spectra of sample S4 with 170 nm VO2 film on $TiO_2$ buffer layer measured at room temperature reveal a typical set of $A_g$ modes characteristic for $VO_2$ M1 monoclinic lattice with positions at 192, 222, 260, 308, 337, 390, 497 and 614 $cm^{-1}$ as well as a trace of a monoclinic M2 mode at 600 $cm^{-1}$ similar to Ref Sample on bare $c\text{-}Al_2O_3$ substrate as shown in Fig. 6 [5]. The absence of modes shift claims that the mechanical stress in lattice of such $VO_2$ films is minor. The detailed information on the detected vibration modes is summarized in Table 3.

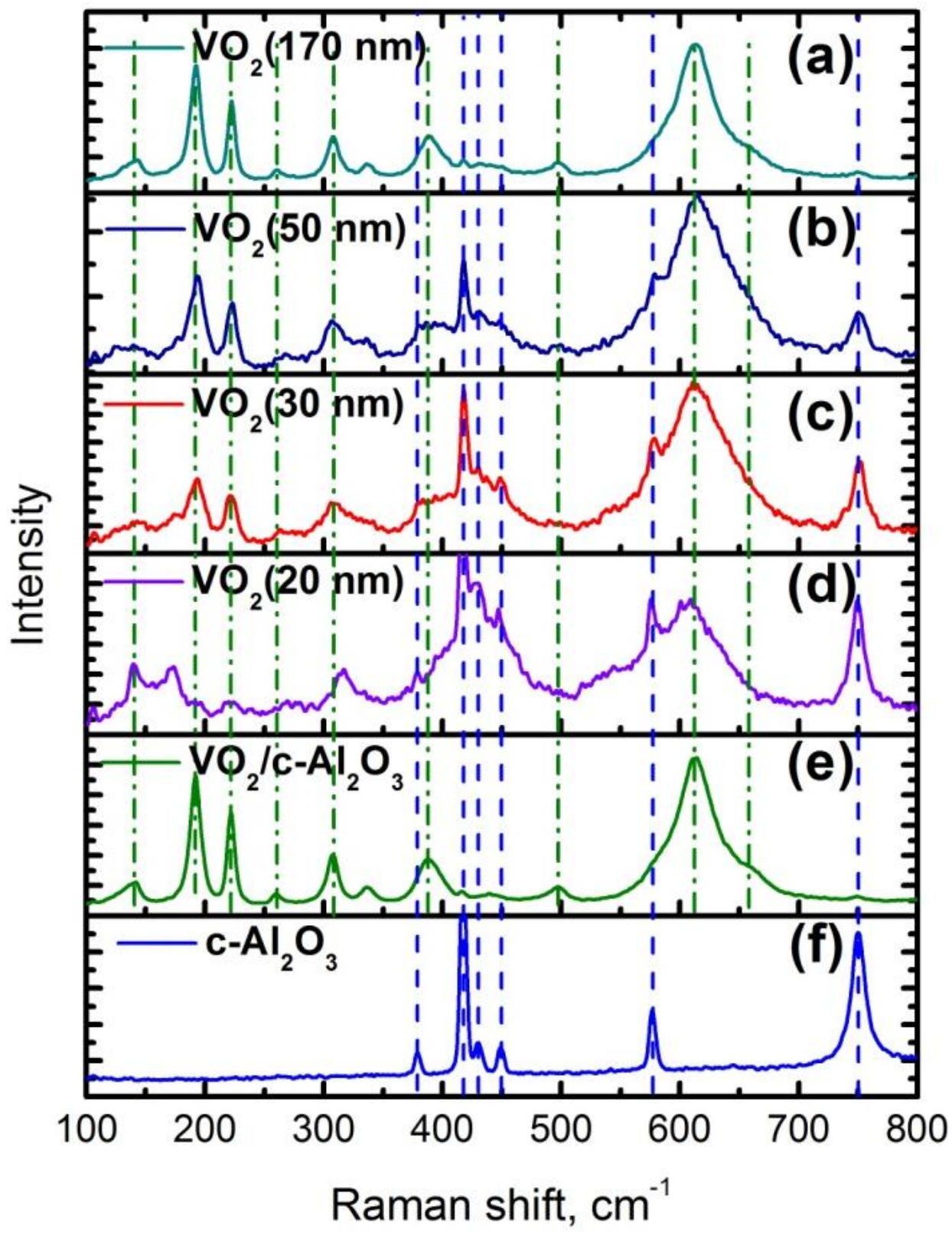

**Fig. 6** Raman spectra of $VO_2$ films with altered thickness: 170 nm (a), 50 nm (b), 30 nm (c), 20 nm (d) deposited on $TiO_2$-buffered c-$Al_2O_3$ substrates. Spectra of $VO_2$ film deposited on bare c-$Al_2O_3$ without buffer layer (e) and of the bare c-$Al_2O_3$ substrate (f).

Remarkable that the positions of $VO_2$ vibration modes are not changed in films when the thickness is decreased to 30 nm. However, in thinner films, in particular when the film thickness is as small as 20 nm, the modes of $VO_2$ reveal significant shifts which evidences that a great mechanical deformation occurs in the crystal lattice. More specifically, one of the two strong $A_g$ lines near 192 $cm^{-1}$ and 222 $cm^{-1}$ is shifted to 173 $cm^{-1}$, while the other is significantly weakened or is completely disappeared when thickness is decreased. The strong $A_g$ mode at 614 $cm^{-1}$ is shifted to 608 $cm^{-1}$. The $A_g$ mode at 308 $cm^{-1}$ which presumably belongs to the M2 phase of $VO_2$ is shifted to 317 $cm^{-1}$. It is notable that the $A_g$ mode at 308 $cm^{-1}$ for 30 nm and 50 nm films is disappeared with increasing temperature and the peak at 317 $cm^{-1}$ which we assign to M2 phase is appeared instead. For a 20 nm film this vibration mode is shifted to 317 $cm^{-1}$ already at room temperature. Also, the $A_g$ mode at 390 $cm^{-1}$ is shifted to 402 $cm^{-1}$ in a 20 nm $VO_2$ film.

Raman spectra recorded at altered temperature as shown in Fig.7. When the semiconductor to metal phase transition is occurred the well pronounced modes corresponding to the monoclinic phase M1 are gradually suppressed or disappear completely and several broad bands of the rutile phase (R) do appear. Typically, the M2 phase in $VO_2$ is observed at ~50 °C, i.e.at temperature just below MIT [21]. It is worth to mention that in present study the thin strained $VO_2$ films do reveal modes of intermediate M2 phase which are detected already at room temperature as shown in Fig.7.

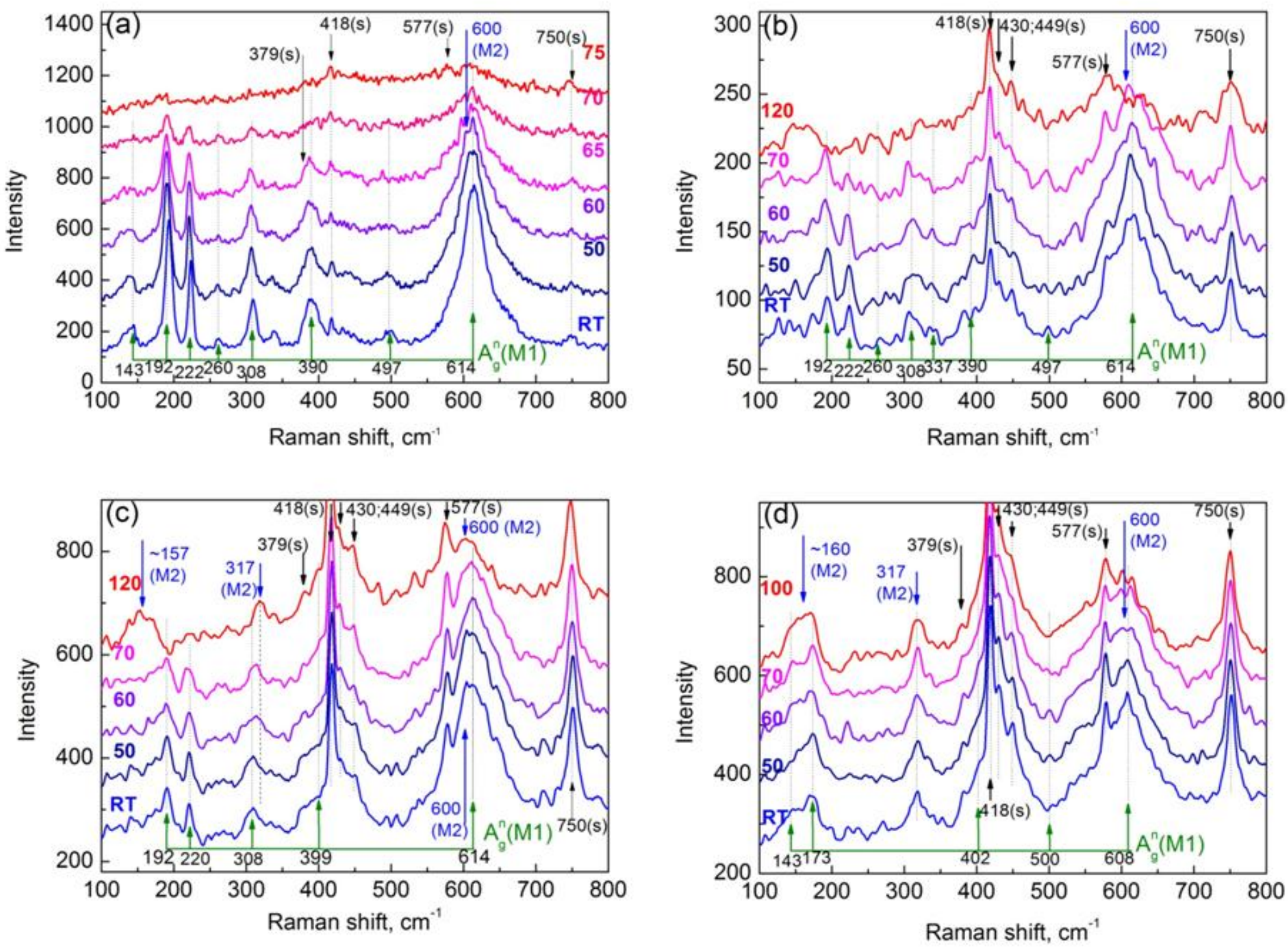

**Fig. 7** Raman spectra of $VO_2$/ $TiO_2$ films of varied thickness as function of temperature. $VO_2$ layer thickness is 170 nm (a), 50 nm (b), 30 nm (c), 20 nm (d).

### 3.5 Influence of mechanical stress in films on MIT

Mechanical strain in $VO_2$ lattice results in altered electrical hysteresis loop as shown in Fig. 8. Reduction of the thickness results to a decreased range of the resistance alteration. Also, MIT temperature is shifted to higher values. It is remarkable that temperature position of the beginning of hysteresis loop is almost independent of $VO_2$ thickness. The resistance of samples in semiconductor state also remains almost unchanged. On the other hand, the resistance of samples in the metallic state is altered significantly with thickness decreasing. More specifically, the metallic state resistance of sample S4 with 20 nm $VO_2$ layer is ~45 times greater at T=90 °C compared to the one of sample S1 with 170 nm film.

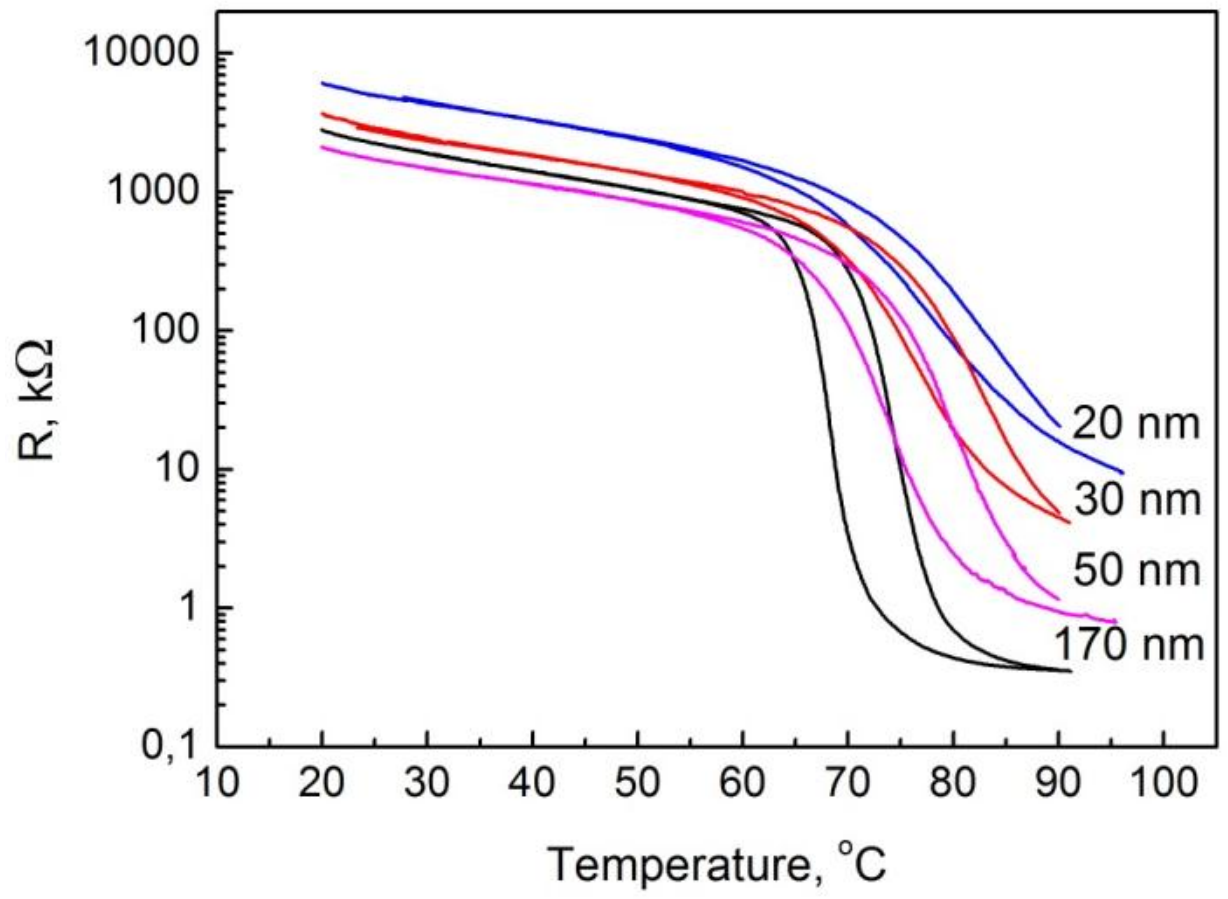

**Fig. 8** MIT triggered resistance change as function of temparature for $VO_2/TiO_2$/c-$Al_2O_3$ samples with different $VO_2$ thickness.

More comprehensive analysis of hysteresis loops is done as shown in Fig. 9 for the temperature dependencies of the derivatives $d(\log_{10}(R))/dT$ that are obtained from data in Fig. 8. The detailed characteristics of MIT hysteresis loops for samples S1-S4 are summarized in Table 4. An average phase transition temperature is increased from 71.5 °C for the Sample S4 with 180 nm $VO_2$ film to 80.7 °C for the Sample S1 with 30nm $VO_2$ layer. Also, a thickness decrease results in increased hysteresis curve abruptness $\Delta T$. Note, that a hysteresis width $\Delta H$ is only slightly altered near 6 °C for all S1-S4 samples. Also, the phase transition temperature in the case of $VO_2/TiO_2/c$-$Al_2O_3$ structures shifts to higher temperatures compared to $VO_2$ /$c$-$Al_2O_3$ structures, while in the case of $VO_2$ films grown on $TiO_2$ single crystal a shift to room temperature was observed with decreasing of the $VO_2$ layer thickness [22].

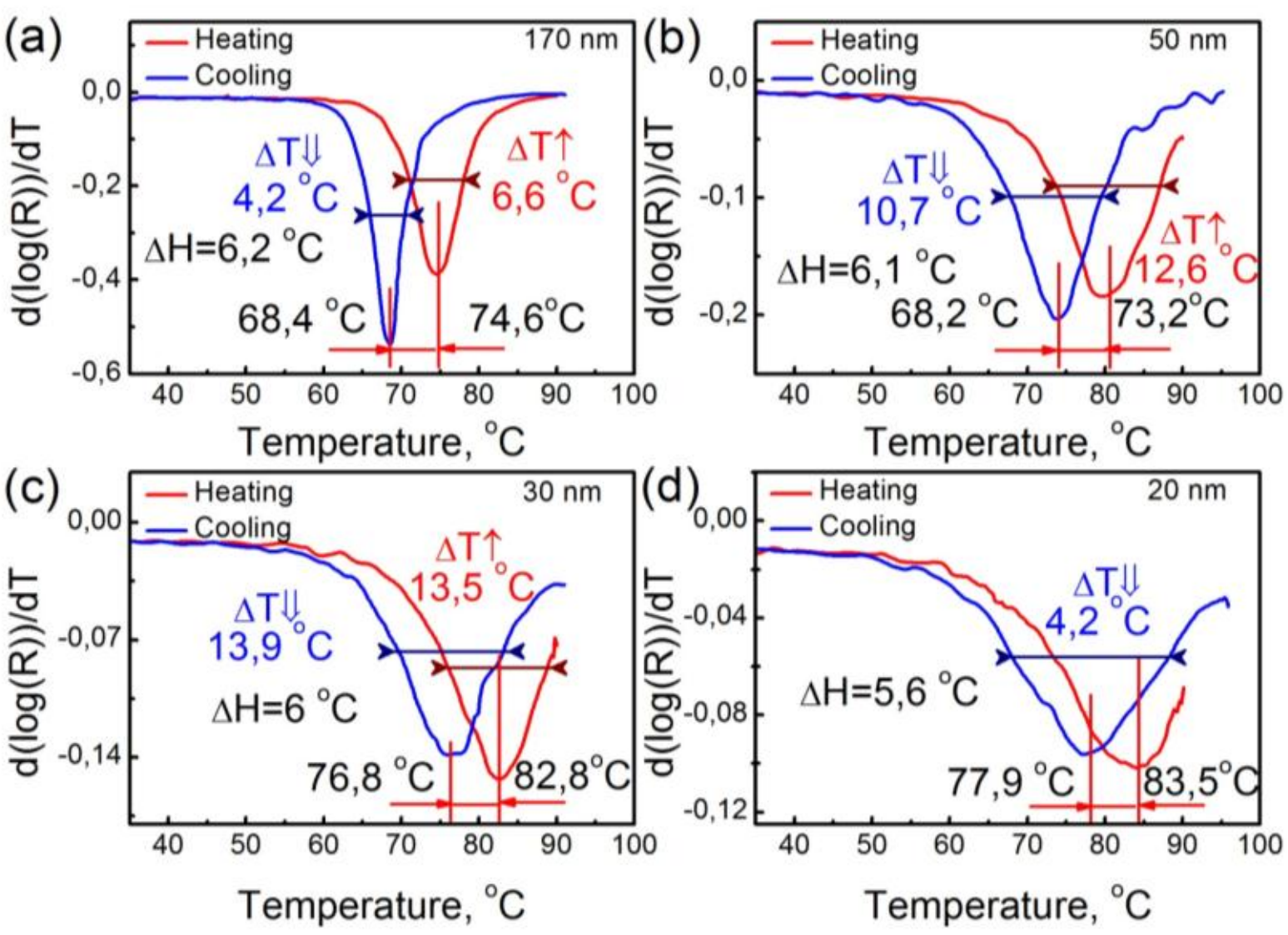

**Fig. 9** Temperature dependence of the derivative $d(\log_{10}(R))/dT$ for samples S4 (a), S3 (b), S2 (c) and S4 (d) with different $VO_2$ layer thickness.

### 3.6 Middle infrared reflection study

The temperature induced increase of $VO_2$ layer conductivity for several orders of magnitude results in dramatic change of electromagnetic waves reflection in the middle infrared spectral range as shown in Fig.10. The dynamics of the reflection change correlates well with temperature dependent evolution of Raman spectra in Fig. 7 and conductivity in Fig. 8. First, the beginning of the isolator to metal phase transition is recognized in spectra already in the range of 60-65 °C (marked as process “A” in Fig. 10) as increase of several IR active vibration mode at ~900 $cm^{-1}$ and several weaker modes in the range of 650-900 $cm^{-1}$ which are observed as increased broad band (reduction of reflection). The phase transition at ~71 °C (marked as “B” in Fig. 10) is well recognized as it results in the growth of a pronounced broad dip centered at ~875 $cm^{-1}$ with reflection reduced to ~1%. At higher temperature range of 71-85 °C the overall increase of reflection is observed in broad spectral range of 650 – 6000 $cm^{-1}$ marked as “C” in Fig. 10 (only part of IR spectrum is presented) and the film becomes conducting. We have experimentally found that the VO2 films fabricated at $2\times10^{-2}$ mbar and a temperature of 550°C do reveal the maximal dynamic range of the middle IR reflection alteration from ~2% to ~64% (at $\lambda$=10 $\mu$m) as shown in Fig. 10. The films deposited at greater or lower oxygen pressure do show smaller MIT transition induced middle IR reflection alteration.

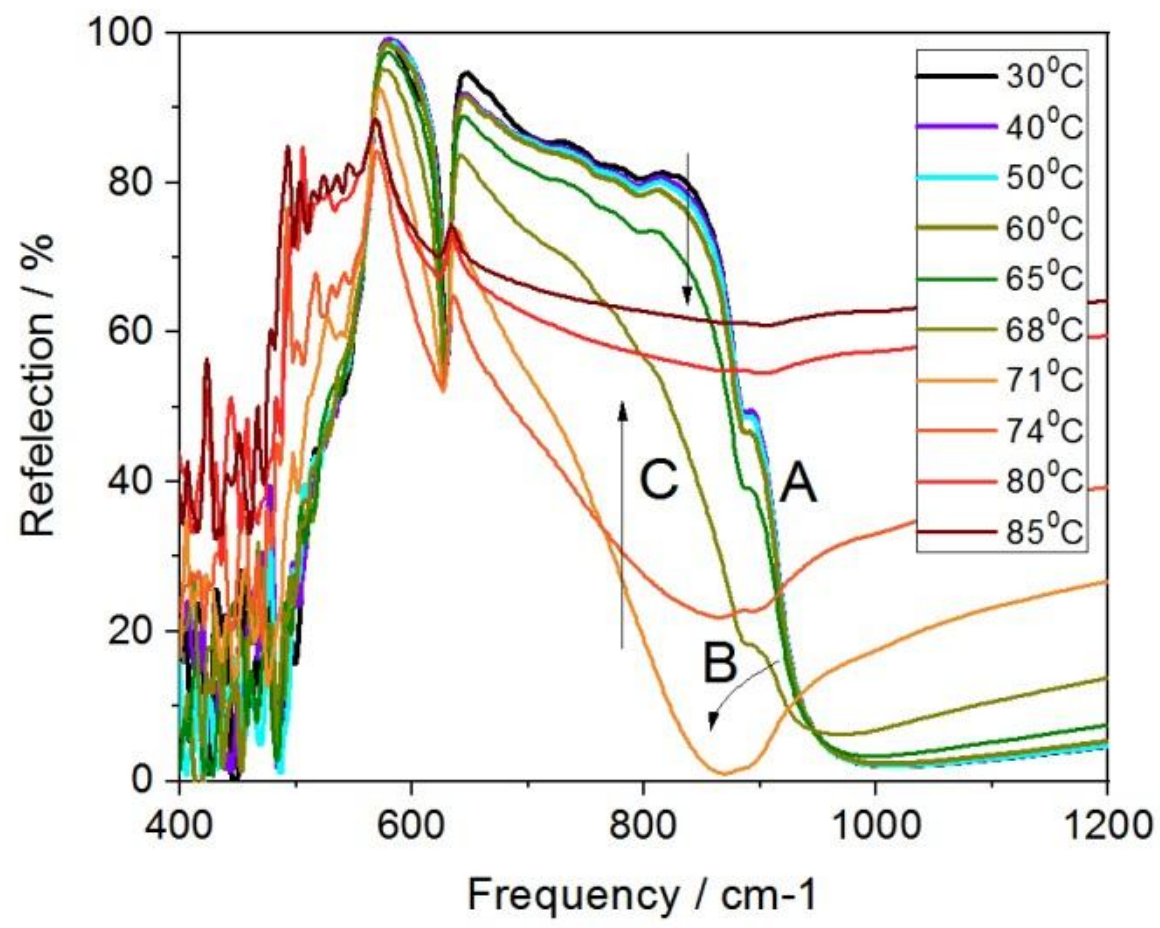

**Fig. 10** Middle infrared reflection spectra of typical $VO_2/TiO_2/c\text{-}Al_2O_3$ 170 nm thick film recorded at altered temperature. $VO_2$ layer was deposited at $2\times10^{-2}$ mbar and temperature of 550°C.

### 3.7 Time resolved electric switching study

Time resolved electric switching characteristics of 5×20 μm $VO_2$ strip fabricated by photolithography from $VO_2/TiO_2/c\text{-}Al_2O_3$ samples S1 and S2 with $VO_2$ thickness of 20 nm and 50 nm are shown in Fig. 11.

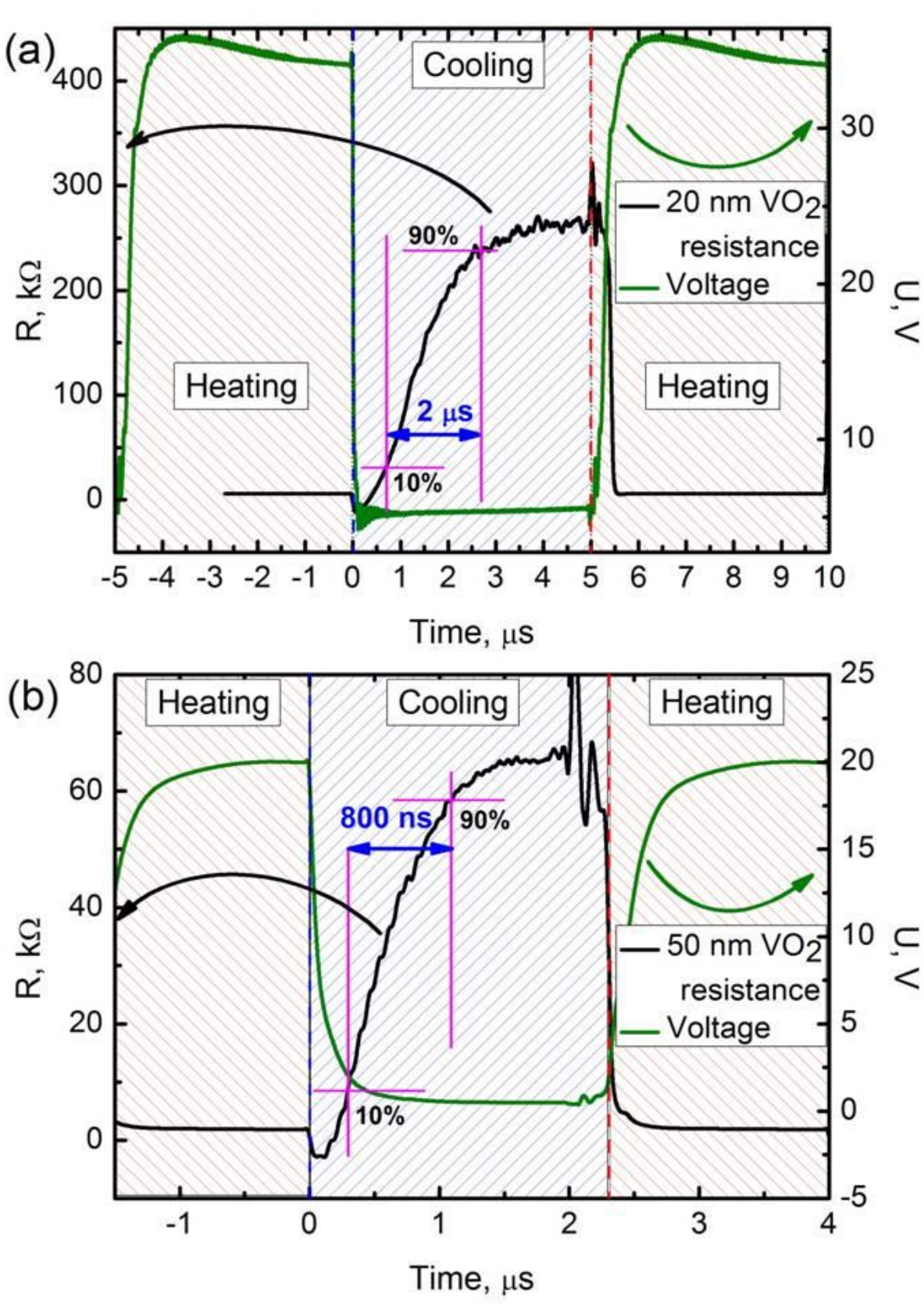

**Fig. 11** Electric induced switching of $VO_2$ films with thickness of 20 nm (a) and 50 nm (b). Voltage applied to the film and corresponding altered film resistance are shown in right and left panels respectively.

Briefly, a pulsed voltage of Vdd=27 V at 500 kHz repetition rate was applied to 50 nm thick $VO_2$. For a 20 nm thick $VO_2$ device a pulsed voltage of Vdd=35 V at 100 kHz repetition rate was applied. The setup used for electric measurement is shown in Fig. 12.

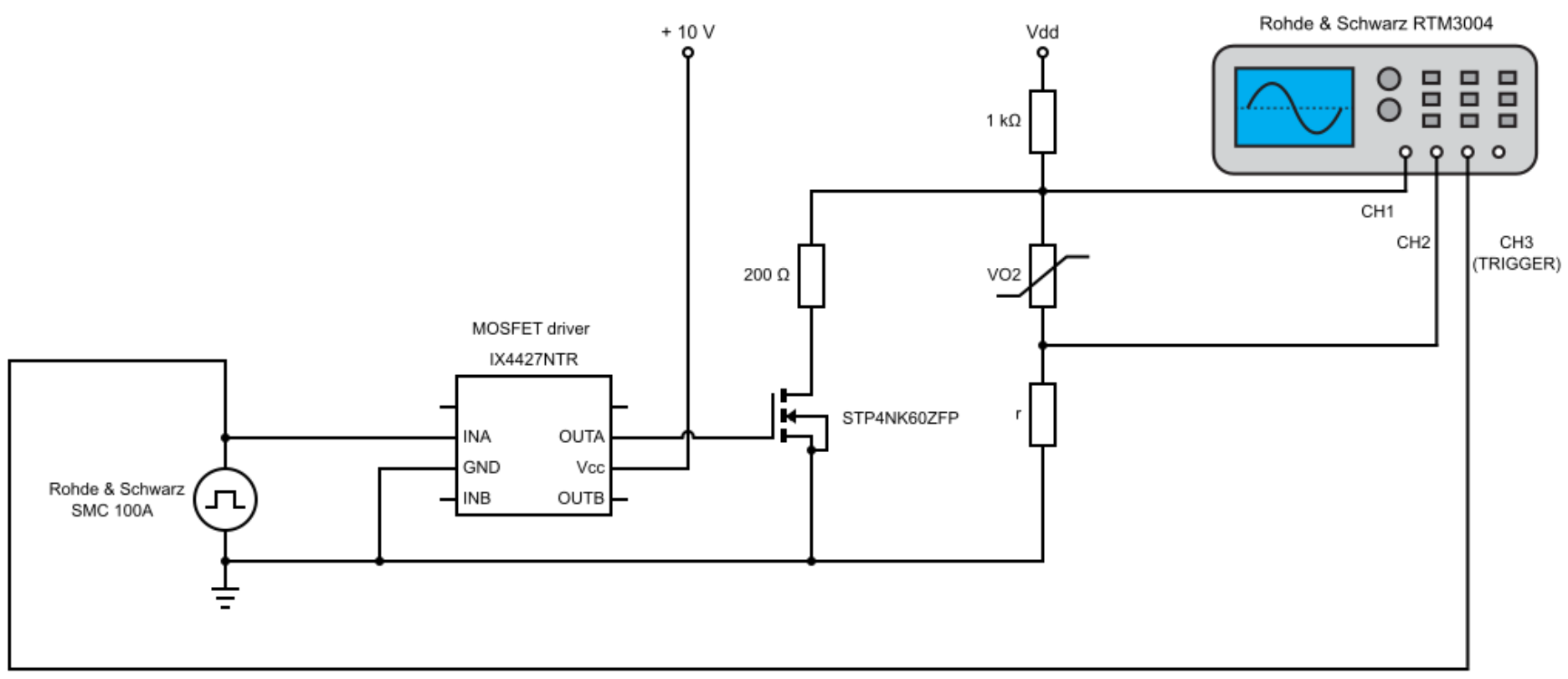

**Fig. 12** Electric circuit scheme for measuring the time resolved switching characteristics of $VO_2/TiO_2$/c-$Al_2O_3$ samples with $VO_2$ films thickness of 20 nm and 50 nm. The limiting resistor (r) was chosen to be 9.1 kOhm and 23.2 kOhm for 50 nm and 20 nm $VO_2$ films correspondingly.

The forward and backward switching times were measured as the time required for resistance to be altered between 10% and 90% as shown in Fig. 11. The resistance switching time from insulator to metallic state due to MIT is about 100 ns for both samples S1 and S2. The backward switching time from metallic to insulator state was measured to be 800 ns and 2 µs for sample S2 and S1 with $VO_2$ thickness of 50 and 20 nm, correspondingly. Indeed, the rate of forward switching is defined by $VO_2$ heating due to Ohmic losses which is proportional to applied current. In case of backward switching the process of heat dissipation is a key factor that limits switching rate. Thus, it can be expected that thinner films should exhibit shorter backwards switching times. However, in spite of thicker $VO_2$ layer the S2 sample showed a faster backward switching rate compared to sample S1. Thus, not only the film thickness but also both the $VO_2$ crystal quality and lattice strain do influence the heat dissipation rate.

The obtained backward switching rates in samples S1 and S2 are slower than value of 252 ns obtained in $VO_2$ films fabricated on more expensive $TiO_2$ single crystal [6]. However, it is not clear weather Lee et. al. measured complete phase transition in $VO_2$ or partial phase transition that is obviously faster. More important, that Lee et al. has also concluded that lattice strain does influence on switching rate. More specifically, significant lattice mismatch between $VO_2$ and $SnO_2$ ~4,2% in [6] results in a pronounced strain release on the interface. The strain relaxation is followed by formation of bulk-like $VO_2$ lattice. The authors claim that forward/ backward switching times in $VO_2/SnO_2/TiO_2$ structure are as short as ~36 ns/~74 ns in such relaxed structure. However, such statement is in a controversy with results obtained by Li et al. [23] where much slower electric switching times of 50 ms and 250 ms reported for $VO_2/V_2O_5$ core-shell nanowire. Indeed, typically a nanowire has much higher crystal quality compared to thin films. However, it was found that nanowire reveal slower switching dynamics compared to its film counterpart [23].

Ultrathin polycrystalline $VO_2$ films prepared on SiN/Si substrates by depositing metallic vanadium layer with molecular beam epitaxy and further annealing in oxygen [7, 8] should be discussed separately. Such films are fabricated for purposes of near IR waveguide modulators. Being polycrystalline they however reveal fast forward and backward switching rates of 2,43 μs and 6,19 μs correspondingly [24]. Again, we could not discriminate weather these times correspond to complete or partial MIT.

It should be underlined that different optical applications do require different degrees of $VO_2$ switching. Even minor alterations of dielectric functions in near IR range which correspond to beginning of phase transition can be already successfully used for fast waveguide modulation. On the other hand, noticeable modulation of THz reflection/absorption is occurred only in the last quarter of the $VO_2$ phase transition just before the metallic state and the switching processes are slower here. The best switching rates obtained to the date are summarized in Table 5.

**Conclusion**

In conclusion, the $TiO_2$-buffered $VO_2$ films show MIT transition at elevated temperature of ~71.5 °C (for 170 nm thick $VO_2$ film) compared to MIT at ~65 °C which is typically observed in $VO_2$/c-$Al_2O_3$ films fabricated in similar conditions. The pulsed laser deposition regime for $VO_2$ films on $TiO_2$-buffered c-$Al_2O_3$ substrates at oxygen pressure of $3\times10^{-2}$ mbar and temperature of 550 °C was found to be optimal to fabricate films with best MIT related resistance alteration characteristics, namely, with most abrupt MIT, greatest resistance alteration ratio $\Delta R_{ratio}$ of $6{,}7\times10^{3}$ as well as a narrow hysteresis loop. The fabricated $VO_2$/$TiO_2$/c-$Al_2O_3$ film structures reveal an epitaxial growth with excellent structural properties: co-direction of the [010]$VO_2$ and [0001]$Al_2O_3$ axes and [010] $VO_2$ axis misorientation smaller than 0.5°, better than for $VO_2$ layer deposited directly on $Al_2O_3$ substrate. The deposition of $VO_2$ on $Al_2O_3$ or on $TiO_2$/$Al_2O_3$ is occurred with distortion of the $VO_2$ unit cell to hexagonal symmetry and the [010] axis of the monoclinic cell of $VO_2$ becomes a sixth order axis.

Properties of $TiO_2$-buffered $VO_2$ films of thickness altered between 170 nm and 20 nm were studied. Composition of $VO_2$ films has revealed only a minor alteration of $VO_2$/$V_2O_5$ phases ratio from 1.6 to 1.8 when the film thickness has been increased from 20 nm to 50 nm and did not change in thicker layers. All the $VO_2$ films thicker than 30 nm did not show any structural strain. However, 20 nm $VO_2$ layers have already reveal a great mechanical deformation of crystal lattice. The MIT temperature is increased from 71.5 °C to 80.7 °C when the thickness of $VO_2$ layer is reduced from 170 nm to 20 nm, however the resistance alteration ratio $\Delta R_{ratio}$ is reduced from $4{,}2\times10^{3}$ to $2.7\times10^{2}$ in thinner films. The backward electric switching times of $VO_2$ films were drastically increased from 800 ns to almost 2 μs with $VO_2$ thickness decrease from 50 nm to 20 nm.

We believe that the obtained results will be highly important when designing $VO_2$-based active IR/THz metasurfaces and all-optical big data processing systems.

**Acknowledgment**

The study was supported by the Ministry of Science and Higher Education of the Russian Federation (state assignment in the field of scientific activity FENW-2023-0014). XRD studies are supported by the Southern Scientific Center of the Russian Academy of Sciences under project No. 122020100294-9. Temperature dependent middle IR reflection spectra study were supported by Russian Science Foundation, grant №24-29-00175 at SFU.

**Table 1** Phase transition temperature $T_{PT}$, abruptness $\Delta T$ , hysteresis width $\Delta H$, minimal resistance $R_{min}$ in metallic state and ratio of the resistance alteration $\Delta R_{ratio}$

| **P($O_2$) / mbar** | **$T_{PT}$ /°C** | **ΔT / °C** | ***$R_{min}$ / kΩ*** | ***ΔH* / °C** | ***$\Delta R_{ratio}$*** |
|---|---|---|---|---|---|
| $1\times10^{-2}$ | 78,6 | 5,1 | 0,04 | 6 | $3,9\times10^{3}$ |
| $2\times10^{-2}$ | 71,45 | 5,7 | 0,35 | 6,1 | $3,1\times10^{3}$ |
| $3\times10^{-2}$ | 70,7 | 3,55 | 0,09 | 5 | $6,7\times10^{3}$ |

*For the transition temperature $T_{PT}$ and abruptness $\Delta T$ the values are given by averaging numbers for heating and cooling curves.*

**Table 2** Binding energies (eV) and relative content (%) of the V2p and O1s components in XPS spectra

| ***Sample*** | **Valence** | **Area*, a.u.** | **$VO_2/V_2O_5$** | **Thickness, nm** |
|---|---|---|---|---|
| S1 | $V^{4+}$<br>$V^{5+}$ | 14,0<br>8,6 | 1,63 | 20 |
| S2 | $V^{4+}$<br>$V^{5+}$ | 17,35<br>10,2 | 1,74 | 30 |
| S3 | $V^{4+}$<br>$V^{5+}$ | 17,7<br>9,85 | 1,8 | 50 |
| S4 | $V^{4+}$<br>$V^{5+}$ | 17,65<br>10,1 | 1,81 | 170 |

* *Area under the curve of the respective peak.*

**Table 3.** Positions of Raman active modes of the $VO_2/TiO_2/c\text{-}Al_2O_3$ samples with various $VO_2$ thickness.

| Laser pulses (thickness) | $VO_2/c\text{-}Al_2O_3$ | 4000 (170 nm) | 1200 (50 nm) | 720 (30 nm) | 480 (20nm) |
|---|---|---|---|---|---|
| Modes | | | Raman shift, $cm^{-1}$ | | |
| $A_g$ $VO_2$ + $B_{1g}$ $TiO_2$ rutile | 142 | 143 | 142 | 143 | 143 |
| $M_2$ $VO_2$ | | | ~157 | ~157 | ~160 |
| $A_g$ $VO_2$ | 192 | 192 | 194 | 192 | 173 |
| $A_g$ $VO_2$ | 222 | 222 | 222 | 222 | 221 |
| $A_g$ $VO_2$ | 260 | 260 | 268 | 268 | |
| $A_g$ $VO_2$ | 308 | 308 | 308 | 308 | 317 |
| $A_g$ $VO_2$ | 337 | 337 | 333 | 337 | 328 |
| *Sapphire* | *379* | *379* | *379* | *379* | *379* |
| $A_g$ $VO_2$ | 390 | 392 | 394 | 399 | 402 |
| *Sapphire* | *417* | *417* | *417* | *417* | *417* |
| *Sapphire* | *430* | | *430* | *430* | *430* |
| *Sapphire* | *449* | | *449* | *449* | *449* |
| $A_g$ $VO_2$ | 497 | 497 | 497 | | 500 |
| *Sapphire* | *577* | *577* | *577* | *577* | *577* |
| $M_2$ $VO_2$ | 600 | 600 | 600 | 600 | 600 |
| $A_g$ $VO_2$ | 614 | 614 | 614 | 614 | 609 |
| *Sapphire* | *750* | *750* | *750* | *750* | *750* |

**Table 4** Temperature of phase transition $T_{PT}$, transition abruptness $\Delta T$ , hysteresis width $\Delta H$, a minimal resistance $R_{min}$ at temperature above phase transition and resistance alteration ratio $R_{ratio}$

| ***Samples*** | **Thickness** | **$T_{PT}$(°C)** | **ΔT (°C)** | ***$R_{min}$* (kΩ)** | ***ΔH* (°C)** | ***$\Delta R_{ratio}$*** |
|---|---|---|---|---|---|---|
| S1 | 20 | 80,7 | ~20 | 15,8 | 5,6 | $2,7\times10^{2}$ |
| S2 | 30 | 79,8 | 13,7 | 4,5 | 6 | $5{,}4\times10^{2}$ |
| S3 | 50 | 77,15 | 11,65 | 0,9 | 6,1 | $2{,}1\times10^{3}$ |
| S4 | 170 | 71,5 | 5,4 | 0,35 | 6,2 | $4{,}2\times10^{3}$ |

*For the transition temperature $T_{PT}$ and transition abruptness $\Delta T$, the given values are averaged data for heating and cooling curves.*

## Table 5.

| Sample | Forward switching time | Backward switching time | Deposition technique | Reference |
|---|---|---|---|---|
| $VO_2/TiO_2$ | 91 ns | 252 ns | PLD | [6] |
| $VO_2/SnO_2/TiO_2$ | 36 ns | 74 ns | PLD | |
| $VO_2$/Si waveguide | 2.43 µs | 6.19 µs | MBE followed by post annealing | [24] |
| $VO_2/V_2O_5$ core-shell single nanobeam | 50 ms | 250 ms | CVD | [23] |
| $VO_2$(50nm)/$SnO_2/TiO_2$ | 100 ns | 800 ns | PLD | present study |